\documentclass[twocolumn,runningheads,natbib]{svjour2}
\smartqed  % flush right qed marks, e.g. at end of proof
\usepackage{graphicx}
\def\ergs {erg\,s$^{-1}$}
\def\ergscm2 {erg\,s$^{-1}$cm$^{-2}$}

\def\rxj {1RXS\,J170849--400910\,}
\def\ea {1E\,2259+586\,}

\def\gtsim{\raisebox{-.5ex}{$\;\stackrel{>}{\sim}\;$}}

\newcommand{\XMM}{{XMM--Newton}\,}
\newcommand{\BSAX}{{\it Beppo}SAX\,}

\journalname{Astrophysics and Space Science}

\begin{document}

\title{X-ray intensity-hardness correlation and deep IR observations of the anomalous X-ray pulsar 1RXS\,J170849--400910\thanks{NR is supported by an NWO Post-Doctoral Fellowship. SZ thanks the Particle Physics and Astronomy Research Coucil, PPARC, for support through an Advanced Fellowship.}}

\titlerunning{1RXS\,J170849--400910: new X-ray and IR observations}        
% if too long for running head

\author{N. Rea         \and
        G. L. Israel   \and
        T. Oosterbroek \and
        S. Campana  \and  
        S. Zane  \and \\
        R. Turolla \and 
        V. Testa \and
        M. M\'endez \and
        L. Stella
}

\authorrunning{Rea et al.} 

\institute{N. Rea \and M. M\'endez \at
 SRON Netherlands Institute for Space Research, \\
 Sorbonnelaan, 2 3584CA, Utrecht, The Netherlands \\
\email{N.Rea@sron.nl}                 
\and
     G.L. Israel
\and V. Testa \and
\and  L. Stella \at
     INAF - Astronomical Observatory of Rome
\and
         T. Oosterboek \at
         Science Payload and Advanced Concepts, ESTEC
\and
     S. Campana \at
     INAF - Astronomical Observatory of Brera
\and
     S. Zane \at
     Mullard Space Science Laboratory, UCL
\and 
    R. Turolla \at
    University of Padua, Physics Department}

\date{Received: date / Accepted: date}

\maketitle

\begin{abstract}
We report here on X--ray and IR observations of the Anomalous X-ray
Pulsar (AXP) \rxj. First, we report on new XMM-Newton, Swift-XRT and
Chandra observations of this AXP, which confirm the
intensity--hardness correlation observed in the long term X-ray
monitoring of this source. These new X-ray observations show that the
AXP flux is rising again, and the spectrum hardening. If the increase
of the source intensity is indeed connected with the glitches and a
possible bursting activity, we expect this source to enter in a
bursting active phase around 2006--2007. Second, we report
on deep IR observations of 1RXS\,J170849--400910, taken with the
VLT-NACO adaptive optics, showing that there are many weak sources
consistent with the AXP position. Neither star A or B, as previously
proposed by different authors, might yet be conclusively recognised as
the IR counterpart of 1RXS\,J170849--400910. Third, using Monte Carlo
simulations, we re-address the calculation of the significance of the
absorption line found in a phase-resolved spectrum of this source, and
interpreted as a resonant scattering cyclotron feature.


\end{abstract}

\section{Introduction}
\label{intro}

AXPs are a small group of neutron stars (NSs) which stand apart from
other known classes of X-ray sources. At the moment there are 7
confirmed AXPs plus 2 candidates. These X--ray pulsars share
many peculiarities: they are all (but one) radio-quiet (Camilo et
al. 2006; Burgay et al. this volume), exhibit X-ray pulsations with
spin periods in a small range of values ($\sim$5--12\,s), they have a
large spin-down rate ($\dot{P}\approx 10^{-13}-10^{-10}$s\,s$^{-1})$,
a rather high X-ray luminosity ($L_{X}\approx 10^{34}-10^{36}$\ergs),
and faint IR counterparts with $K_s\sim20-22$\,magnitudes (for a
recent review see Woods \& Thompson 2004 and Kaspi 2006 in this
volume). The nature of their X-ray emission was intriguing all
along. In fact, it is too high to be produced by the loss of
rotational energy alone, but on the other hand, no hints for a
companion star were found, neither through deep observations at other
wavelength, nor timing the X-ray pulsations with the hope of finding
Doppler shifts (Israel et al. 2003a; Mereghetti, Israel \& Stella
1998).

%%%%%%%%%%%%%%%%% Table 1 %%%%%%%%%%%%%%%%%%%%%%%%%%%%

\begin{table*} 
\caption{ Best fit values of the spectral parameters obtained for 
about ten years X-ray monitoring of \rxj. Fluxes (and percentages of
fluxes) are unabsorbed, in units of
$10^{-10}$\,erg\,s$^{-1}$\,cm$^{-2}$ and in the 0.5--10\,keV energy
range. The $N_{H}$ was fixed at the XMM-Newton value of
$1.36\times10^{22}$\,cm$^{-2}$ for all the observations. See text for
details. Errors are at 90\% confidence level.}

\centering
\label{tab:1}

\begin{tabular}{lccccccc}
\hline\noalign{\smallskip}

& ASCA\,1996 & SAX\,1999 & SAX\,2001 & Chandra\,2002 & XMM\,2003 & Chandra\,2004 & Swift\,2005 \\
\hline\noalign{\smallskip}

kT (keV) & 0.41$^{+0.01}_{-0.01}$ & 0.465$^{+0.002}_{-0.017}$ & 0.424$^{+0.003}_{-0.006}$ & 0.475$^{+0.0}_{-0.02}$ & 0.456$^{+0.007}_{-0.004}$ & 0.43$^{+0.01}_{-0.01}$ & 0.430$^{+0.015}_{-0.017}$ \\

Gamma  & 2.51$^{+0.11}_{-0.11}$ & 2.65$^{+0.08}_{-0.03}$ &  2.45$^{+0.04}_{-0.03}$ & 2.47$^{+0.11}_{-0.1}$ &2.792$^{+0.008}_{-0.012}$ & 2.77$^{+0.03}_{-0.08}$ &2.62$^{+0.04}_{-0.02}$ \\

Flux & 1.5$^{+0.1}_{-0.08}$ & 1.23$^{+0.04}_{-0.05}$ & 1.30$^{+0.013}_{-0.015}$  & 1.06$^{+0.02}_{-0.02}$ &0.87$^{+0.004}_{-0.002}$ &1.30$^{+0.04}_{-0.06}$ & 1.43$^{+0.01}_{-0.04}$ \\
PL Flux (\%)& $82\pm9$ & $73\pm4$ & $69\pm3$ & $74\pm3$ & $84\pm1$  & $83\pm 3$  & $71\pm3$ \\
$\chi^2_{\nu}$ (d.o.f.) &  1.05 (71) & 1.07 (148) & 1.19 (215)  & 0.93 (430) & 1.14 (221) & 0.95 (147)  & 1.11 (182) \\
\tableheadseprule\noalign{\smallskip}
\noalign{\smallskip}\hline
\end{tabular}
\end{table*}

%%%%%%%%%%%%%%%%%%%%%%%%%%%%%%%%%%%%%%%%%%%%%%%%%%%%%%%%%%%%%%%%%%%%%%%%%%%%

At present, the model which is most successful in explaining the
peculiar observational properties of AXPs is the ``magnetar''
model. In this scenario AXPs are thought to be isolated NSs endowed
with ultra-high magnetic fields ($B\sim 10^{14}-10^{15}$\,Gauss) and
their X-ray emission powered by magnetic field decay (Duncan \&
Thompson 1992; Thompson \& Duncan 1993, 1996). This idea is
strongly supported by the estimate of AXPs' magnetic field through the
classical dipole braking formula,
$B\sim3.2\times10^{19}\sqrt{P\dot{P}}$~Gauss, which gives in all cases
values above the electron critical magnetic field ($B_{QED}\sim
4.4\times10^{13}$~Gauss).

Alternative scenarios, invoking accretion from a fossil disk remnant
of the supernova explosion (van Paradijs, Taam \& van den Heuvel 1995;
Chatterjee, Hernquist \& Narayan 2000; Alpar 2001), are still open
possibilities although encounter increasing difficulties in explaining
the data.

%AXPs were revealed thanks to their persistent emission in the soft
%X-rays, at variance with the Soft Gamma-Ray Repeaters (SGRs), another
%class of possible ``magnetars'', which were discovered thanks to their
%bursting activity in the X/$\gamma$-rays, only later on discovered
%also from AXPs (Gavriil, Kaspi \& Woods 2002).

\rxj\ was discovered with ROSAT (Voges et al. 1996) and later on a
$\sim$\,11\,s modulation was found in its X-ray flux with ASCA
(Sugizaki et al. 1997).  Early measurements suggested that it was a
fairly stable rotator with a spin period derivative of $\sim
1.9\times10^{-11}$s\,s$^{-1}$ (Israel et al. 1999). However, in the
last four years the source experienced two glitches, with different
post-glitch recoveries (Kaspi, Lackey \& Chakrabarty 2000, Dall'Osso
et al. 2003, Kaspi \& Gavriil 2003). Searches for optical/IR
counterparts ruled out the presence of a massive companion (Israel et
al. 1999). Very recently, two different objects were proposed by
different groups, as being \rxj\, IR counterpart, and there still is
an open debate on which one is the AXP counterpart (Israel et
al. 2003b; Safi-Harb \& West 2005; Durant \& van Kerkwijk 2006). A
diffuse ($\sim 8^{\prime}$) radio emission at 1.4 GHz was recently
reported, possibly associated with the supernova remnant G346.5--0.1
(Gaensler et al. 2000).

Pulse phase spectroscopy analysis of two \BSAX\, observations of
\rxj\, (Israel et al. 2001; Rea et al. 2003) revealed i) a large
spectral variability with the spin-phase, ii) a strong energy
dependence of the pulse profile shape, and iii) shifts in the pulse
phase between the low and the high energy profiles. High variability
of the pulse shape with energy is now detected at higher energies, up
to $\sim$220~keV (Kuiper et al. 2006).

By analysing a \BSAX\,observation taken in 2001 (the longest pointing
ever performed on this source), Rea et al. (2003) reported the
presence of an absorption line at $\sim 8$ keV in a phase-resolved
spectrum. Interpreting the feature as a cyclotron line due to resonant
scattering yields a neutron star magnetic field of either
$9.2\times10^{11}$\,G or $1.6\times10^{15}$\,G, in the case of
electron or proton scattering, respectively.

In \S\ref{observations} we report on new \rxj\, XMM-Newton, Chandra
and Swift observations which were used together with the previous
ROSAT, ASCA and \BSAX\ observations to monitor the X-ray spectrum and
flux of the AXP. Then we report on deep infrared VLT-NACO observations
of this AXP. In \S\ref{cyclo} we carefully re-address the calculation
of the significance of the absorption line found around 8\,keV during
a long \BSAX\, observation (Rea et al.~2003), we then summarise and
discuss all the results in \S\ref{diss}.

\section{Observations and Results}
\label{observations}

In this section we report on the observations and the data analysis of
the four new X--ray and the IR observations. The results on the X--ray
timing analysis are reported below in the text while the X--ray
spectral parameters may be found in Tab.\,1. All the X-ray spectra
were fit by an absorbed blackbody plus a power--law component
(but see also Rea, Zane, Lyutikov \& Turolla in this volume for a
different spectral modelling). Giving the very high statistics we have
in the XMM--Newton observation, and assuming the absorption does not
vary along the line of sight, we fixed for all the spectra the
absorption at the XMM--Newton value ({\tt phabs} XSPEC model: $N_{H} =
(1.36\pm0.04)\times10^{22}$\,cm$^{-2}$; abundances from Anders \&
Grevesse 1989)

\subsection{XMM--Newton}

\rxj\, was observed with \XMM\, between 2003 August 28th and 29th,
for $\sim50$~ks. The MOS cameras were operated in Prime Partial Window
Mode, while the PN camera was in Prime Small Window Mode, all with the
medium optical photons blocking filter.  Since a higher background
affected the last $\sim$10\,ks of the observation, we used only the
data during intervals in which the count rate above 10~keV was less
than 0.35 counts s$^{-1}$.  The source events and spectra were
extracted within a circular region of 27$^{\prime\prime}$ centred on
the peak of the point spread function of the source. This non standard
radius was used because the source was located near the edge of the
chip. The background was obtained from a source-free region of
27$^{\prime\prime}$. In order to determine the spin period of \rxj\,
we barycentered the events arrival times and obtained, through a
phasefitting technique, a best spin period of
$P_{s}=11.00170\pm0.00004$\,s (all errors are at the 90\,\% confidence
level). We found that the pulsed fraction of the X-ray signal (defined
as the amplitude of the best-fitting sine wave divided by the,
background corrected, constant level of the emission) is
energy-dependent, and it varies from $39.0\pm0.5$\% in the
0.5--2.0\,keV range to $29\pm1.5$\% in the 6.0--10.0\,keV range. These
values are consistent with those reported for the pre-glitches
\BSAX\,observation (Israel et al. 2001) while both are larger than
those reported for the post-glitches \BSAX\ observation (Rea et
al. 2003). Detailed results for this observation are reported in Rea
et al.~(2005a).

%%%%%%%%%%%%%%%%%%%%%%%%%%%%%%%%%%%%%%%%%%%%%%%%%%%%%%%%%%%%%%%%%%%%%%%
\begin{figure}[t]
\centering
\includegraphics[height=0.35\textwidth,width=0.35\textwidth]{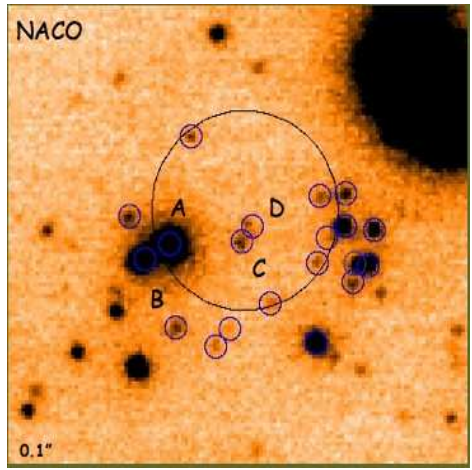}
\caption{VLT-NACO image in the $K_s$ band of the field of \rxj. The Chandra $0.8^{\prime\prime}$ 90\% error circle is over-plotted (Israel et al.~2003b). We marked all the faint sources being consistent at 3$\sigma$ with the AXP position. Note that our astrometry (following Israel et al.~2003b) is slightly different from the one reported by Durant \& van Kerkwijk~(2006), this is due to a different catalogue used for the astrometry.}
\label{naco_rxs}       % Give a unique label
\end{figure}
%%%%%%%%%%%%%%%%%%%%%%%%%%%%%%%%%%%%%%%%%%%%%%%%%%%%%%%%%%%%%%%%%%%%%%%

\subsection{Chandra}

\rxj\, was observed by the {\em Chandra} Advanced CCD 
Imaging Spectrometer (ACIS), first for $\sim$30\,ks with the High
Energy Transmission Grating Spectrometer (HETGS) on 2002 September
9th, then for $\sim$30\,ks in Continuous Clocking (CC) mode on 2004
July 3. For a more detailed description of the instruments and on the
data processing we defer to the {\it Chandra\/} X-ray Center (CXC)
documents\footnote{http://asc.harvard.edu/udocs/docs/docs.html;
http://asc.harvard.edu/ciao/}. Detailed results for these two
observations are reported in Rea et al.~(2005a) and Campana et
al.~(2006).

\subsubsection{High Energy Transmission Grating Spectrometer}

The High Energy Transmission Grating Spectrometer (HETGS) employs two
sets of transmission gratings: the Medium Energy Gratings (MEGs) with
a range of 2.5--31~\AA\/ (0.4--5.0~keV) and the High Energy Gratings
(HEGs) with a range of 1.2--15~\AA\/ (0.8--10.0~keV).  The HETGS
spectra were imaged by ACIS-S, an array of 6 CCD detectors normally
read-out every 3.2\,s.  The HETGS/ACIS-S combination provides an
undispersed (zeroth order) image and dispersed spectra from the
gratings.  The various orders overlap and are sorted using the
intrinsic energy resolution of the ACIS CCDs: $\Delta\lambda=$
0.012~\AA\/ for the HEG and 0.023~\AA\/ for the MEG.

The MEG and HEG first order count rate were only 0.5 and
0.2~cts~s$^{-1}$, we therefore did not expect the dispersed spectrum
to be affected by pileup, while the zeroth-order image was not used in
our spectral analysis because highly affected by photon pileup. We
used the standard CIAO tools to create detector response files for the
MEG and HEG $+1$ and $-1$ order spectra.  These were combined when the
$+$/$-$ order spectra were added for the HEG and MEG separately. We
binned the data at 0.08~\AA\/ with a minimum of 30 counts per bin.  To
look for high-resolution spectral features, the data were binned at
0.015~\AA\/ for the HEG and 0.03~\AA\/ for the MEG.  We also created
background files for the HEG and MEG spectra using the standard CIAO
tools.

\subsubsection{Continuous Clocking}

In order to avoid pile-up effects, in this second observation the source
was observed in the Continuous Clocking (CC) mode (CC33\_FAINT; time
resolution 2.85\,ms). The source was positioned in the
back-illuminated ACIS-S3 CCD on the nominal target position.  A
detailed description on the analysis procedures, such as extraction
regions, corrections and filtering applied to the source events and
spectra can be found in Rea et al.~(2005b).

In order to perform a timing analysis we corrected the events arrival
times for the barycenter of the solar system (with the CIAO {\tt
axbary} tool) using the provided ephemeris. We used for the timing
analysis only the events in the 0.3--8\,keV energy range and the
standard {\it Xronos} tools (version 5.19). One fundamental peak plus
one harmonic were present in the power-spectrum. A period of
$11.00223\pm0.00005$\,s was detected referred to MJD 53189. The pulse
profile did not change with respect to the previous detection and the
0.3--8\,keV PF is $35.4\pm0.5$\%.  Being the CC mode not yet
spectrally calibrated, the Timed Exposure (TE) mode response matrices
(rmf) and ancillary files (arf) are generally used for the spectral
analysis. We defer to Rea et al.~(2005b) for a detailed description on
the extraction procedures of the spectral matrices.

\subsection{Swift}

\rxj was observed with the {\em Swift} satellite a few times 
between 2005 January 29th and March 29th, being a calibration source
for the timing accuracy and for the wings of the Point Spread Function
of the X--Ray Telescope (XRT).  Here we focus on data taken in Window
Timing (WT) and Photon Counting (PC) mode longer than 1 ks.  We
extracted data from two WT observations. The extraction region is
computed automatically by the analysis software and is a box 40 pixels
along the WT strip, centred on source, encompassing $\sim 98\%$ of the
Point Spread Function in this observing mode. We extracted photons
from PC data from an annular region (3 pixels inner radius, 30 pixel
outer radius) in order to avoid pile-up contamination. We consider
standard grades 0--2 in WT and 0--12 in PC modes. Background spectra
were taken from close-by regions free of sources. The photon arrival
times were corrected to the Solar system barycenter. A period search
led to a clear detection of the neutron star spin period at
$P=11.0027\pm0.0003$\,s, derived with a phase fitting techniques. This
period is consistent with the extrapolation from known ephemerides at
a constant period derivative (Kaspi \& Gavriil 2003; Dall'Osso et
al. 2003). We found a PF of $31\pm2\%$, $39\pm3\%$, $29\pm4\%$ and
$35\pm7\%$ in the 0.2--10 keV, 0.2--2 keV, 2--4 keV and 4--10 keV
energy bands, respectively. Detailed results are reported in Campana
et al.~(2006).

\subsection{IR observation: VLT--NACO}

A deep observation of the \rxj\, field, was taken on 2003 June 20th
from the Very Large Telescope using the NAOS--CONICA adaptive
optics. We defer to Israel et al.~2004 for details in the data
reduction. In Fig.\ref{naco_rxs} we present the $K_s$ band field
around the Chandra $0.8^{\prime\prime}$ position of \rxj (Israel et
al.~2004). Besides sources A and B proposed by Israel et al.~(2003b),
Safi--Harb \& West~(2005) and Durant \& van Kerkwijk (2006), as the
possible IR counterparts to this AXP, many further faint sources
(e.g. Star C and Star D) were detected strongly consistent with the
Chandra uncertainty region. Unfortunately, $H$ band images were
obtained under poor sky conditions, and thus the $H-Ks$ colour could
not be determined for these other faint objects.  Nonetheless, we note
that the $Ks$ magnitudes of e.g. stars C and D ($20.3\pm0.2$ and
$21.7\pm0.3$, respectively) are in better agreement with the IR
magnitudes typical of AXPs, than star A and B previously proposed
($17.61\pm0.07$ and $18.78\pm0.05$, respectively). Deep images in the
$L'$ band were also obtained, but no object was detected within the
Chandra uncertainty region at a limiting magnitude of
$L^{\prime}\sim17.8$, the deepest limit ever obtained for an AXPs in
this band. Detailed results and discussion will be reported in Israel
et al.~(2006, in prep.).

%%%%%%%%%%%%%%%%%%%%%%%%%%%%%%%%%%%%%%%%%%%%%%%%%%%%%%%%%%%%%%%%%%%%%%%
\begin{figure}[t]
\centering
\includegraphics[height=0.42\textwidth,angle=-90,width=0.43\textwidth]{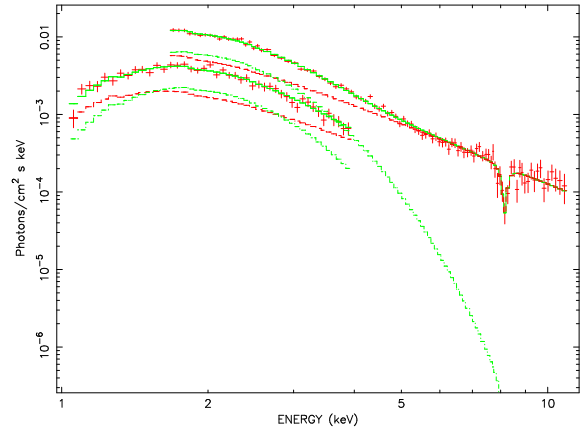}
\caption{Unfolded phase-resolved spectrum presenting the absorption line discovered in \rxj (Rea et al. 2003; Rea et al. 2004).}
\label{line}       % Give a unique label
\end{figure}
%%%%%%%%%%%%%%%%%%%%%%%%%%%%%%%%%%%%%%%%%%%%%%%%%%%%%%%%%%%%%%%%%%%%%%%

\section{On the absorption line at 8.1\,keV}
\label{cyclo}

During the post glitches \BSAX\, observation in 2001, evidence for an
absorption feature was found (see Fig.\ref{line}). This feature was
not detected during the XMM\, observation 3 years later, the only
observation by now with a comparable statistics. The \XMM\, upper
limit for the line depth is 0.15 at 95\% confidence level, which
compared with the value found by Rea et al. (2003; 0.8$\pm$0.4 at 90\%
confidence level), leave only a very small chance that the two
measurements are consistent.

%%%%%%%%%%%%%%%%%%%%%%%%%%%%%%%%%%%%%%%%%%%%%%%%%%%%%%%%%%%%%%%%%%%%%%%
\begin{figure*}[t]
\begin{center}
\hbox{
\includegraphics[height=0.5\textwidth,angle=-90,width=0.5\textwidth]{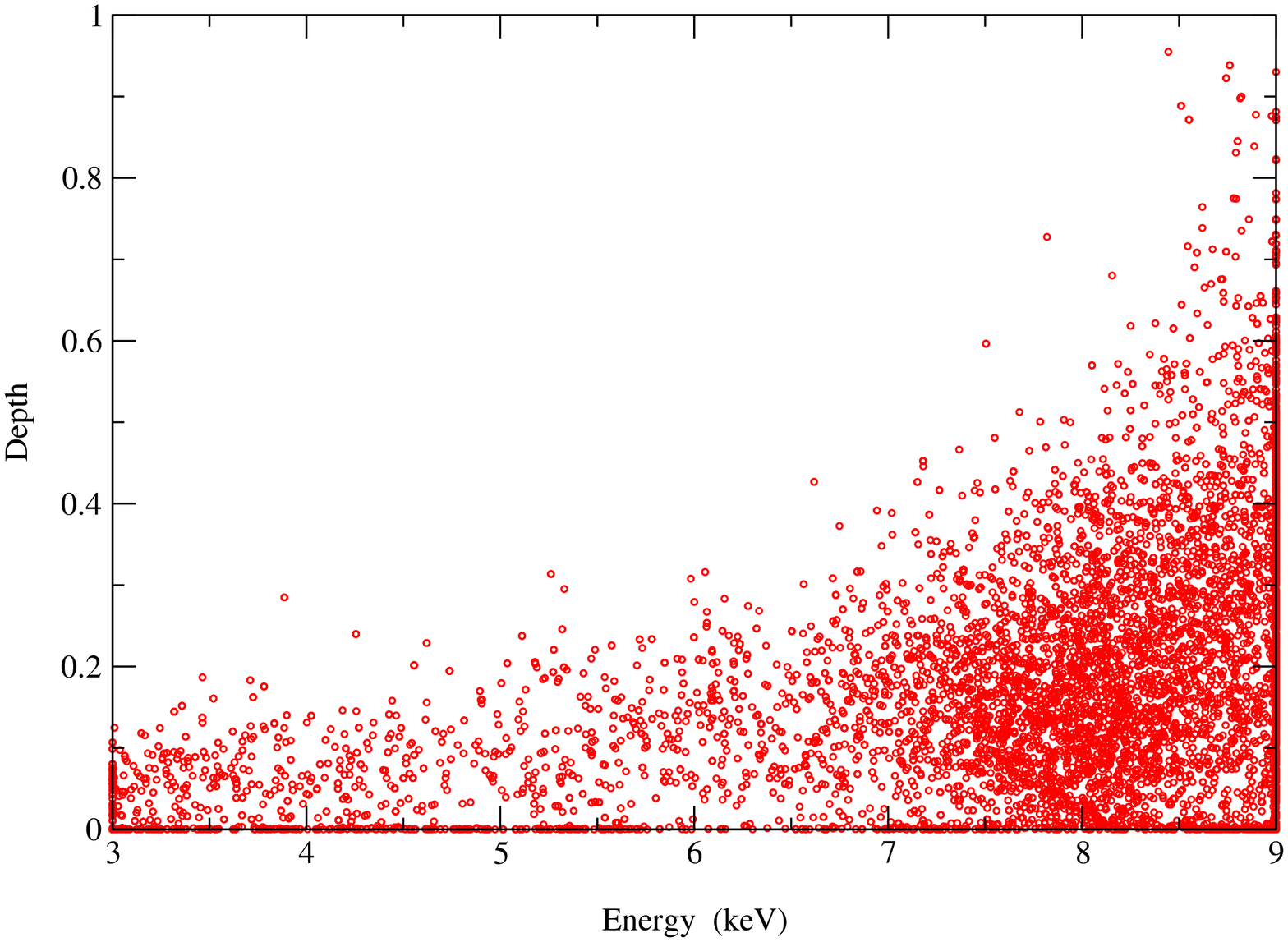}
\includegraphics[height=0.5\textwidth,angle=-90,width=0.5\textwidth]{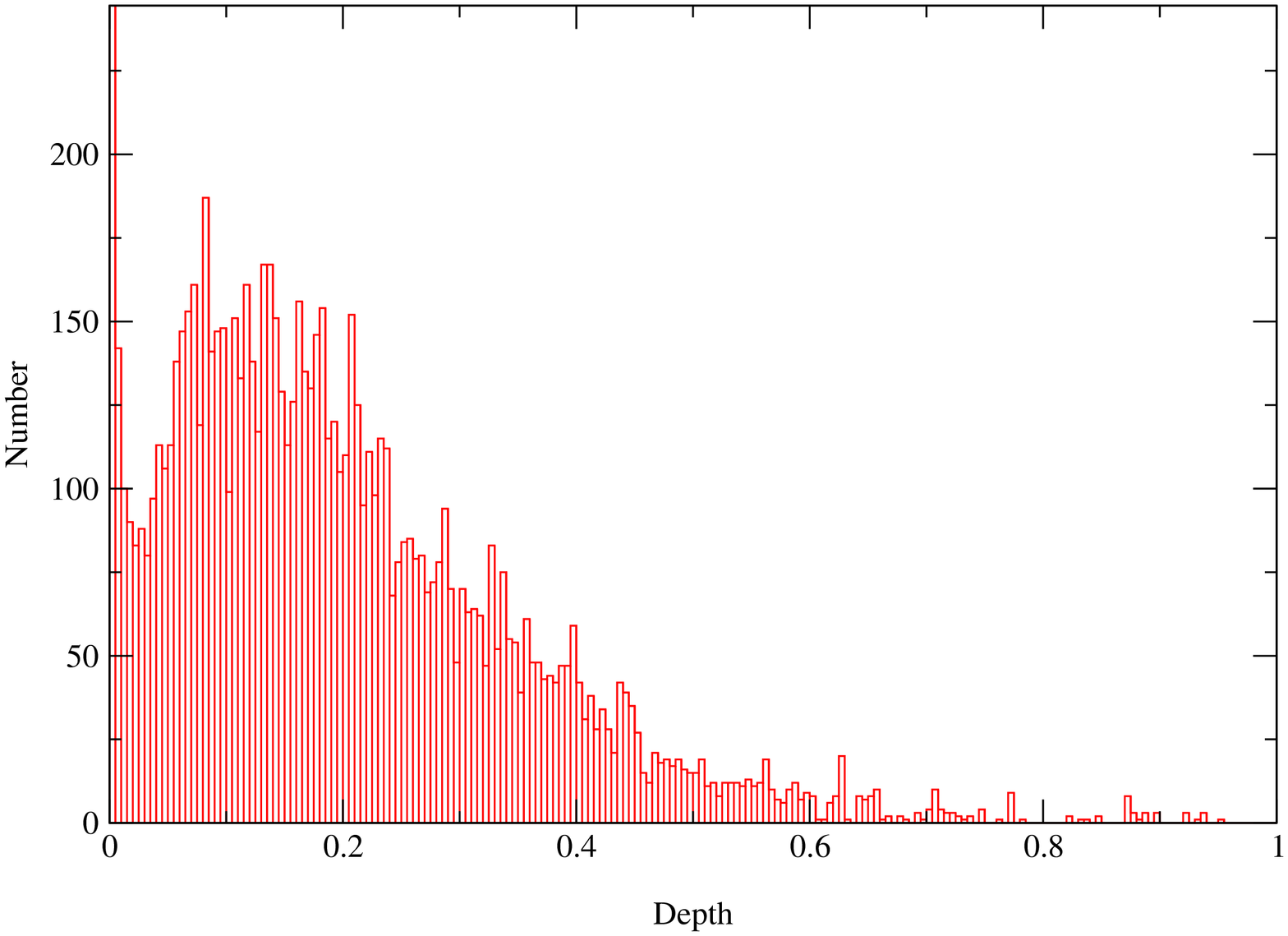}
}

\caption{Results of the Monte Carlo simulation of $10^4$ spectra. {\em Left Panel}: Depth versus energy of the detected lines. {\em Right Panel}: Number of spectra for which a line at 8.1\,keV was detected, as a function of the line depth.}
\end{center}
\label{montecarlo}       % Give a unique label
\end{figure*}

%%%%%%%%%%%%%%%%%%%%%%%%%%%%%%%%%%%%%%%%%%%%%%%%%%%%%%%%%%%%%%%%%%%%%%%

We then undertook a careful re-analysis of the \BSAX\, data. This
re-analysis resulted in the finding that the phases at which the
absorption line was strongest were given incorrectly in the published
version of Rea et al. (2003), as we noticed earlier: in particular,
the line is strongest close to the pulse minimum in the 0.1-2\,keV
band (or the pulse maximum in the 6-10\,keV band).  We nevertheless
found that the reported estimate of the significance is sound and not
much influenced by different choices in the background subtraction
(annular regions, circular regions far from the source or using blank
field files) or by different extraction regions for the source. The
re-analysis of the \BSAX\, data made varying the extraction radius,
the criterion for the background subtraction and the spectral binning
factor, results in basically the same line properties, which
strengthen our confidence in the robustness of the result.

Using an F-test method and taking into account the six trials we made
in the phase resolved spectra, we derive a confidence level for the
absorption line of $\sim 4\sigma$.  Note that even if we take into
account all the possible energies at which the feature could lie in
the LECS plus MECS energy range, the confidence level is still $\gtsim
3.5\sigma$. However, despite being the most common method in
astrophysics to derive the significance of the emission or absorption
spectral features, Protassov et al.~(2002) pointed out that the F-test
may be inappropriate in these circumstances, leading sometimes to
incorrect significance estimates.

Following the recipe of Protassov et al.~(2002), in order to further
investigate on this significance issue, we ran a Monte Carlo
simulation of $10^4$ spectra fixing only the continuum model
(parameters reported in Rea et al. 2003) and the same number of
photons of the phase resolved spectrum which showed the line in the
2001 \BSAX\, observation. The results of the simulation is shown in
Fig.\ref{montecarlo}: in the left panel each red circle represents one
of the $10^{4}$ simulated spectra for which a line was detected, here
we plot the depth of the lines as a function of the line energy. On
the other hand, in the right panel we report the the number of
spectra, among the $10^{4}$ simulated spectra, for which the
statistical fluctuation reported the presence of a line at 8.1\,keV,
as the function of the line depth. From this simulation we found 32
spectra with depth $>$0.8 in $10^4$ points. We can then reliably say
that the probability of the line being a fluctuation is $<$0.32\%. In
summary, we confirm the detection of the 8.1\,keV absorption line in
the \BSAX\, data made by Rea et al. (2003) at 99.68\% confidence level
(see Fig.\ref{montecarlo}). Note that the non homogeneous coverage of
red circles over the entire spectral range mirrors the energy
dependency of the \BSAX\, spectral matrices.

%%%%%%%%%%%%%%%%%%%%%%%%%%%%%%%%%%%%%%%%%%%%%%%%%%%%%%%%%%%%%%%%%%%%%%%
\begin{figure*}[t]
\centering
\includegraphics[height=0.3\textwidth,angle=-90,width=0.6\textwidth]{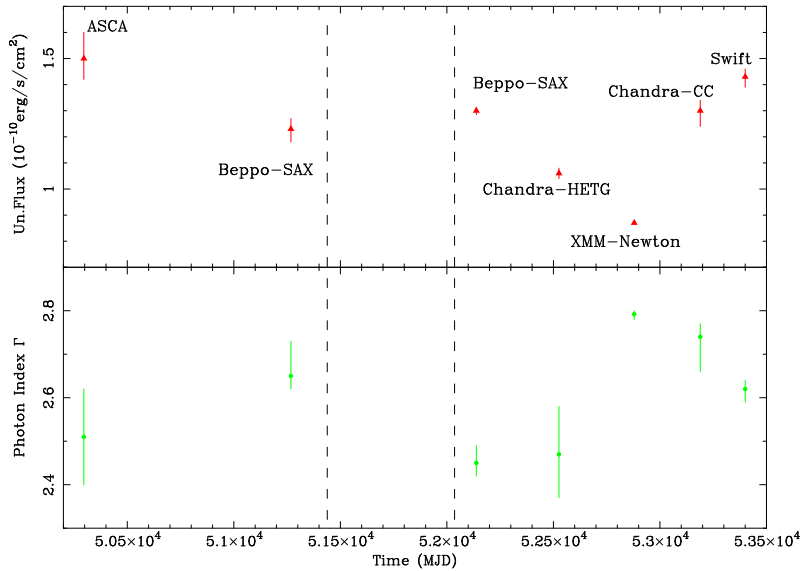}
\caption{Correlated photon index and X-ray flux changes with time. 
Dashed line represents the two glitch epochs. All
fluxes are unabsorbed and calculated in the 0.5-10\,keV energy
band keeping fixed the $N_{H} = 1.36\times10^{22}$\,cm$^{-2}$.}
\label{plot_all}       % Give a unique label
\end{figure*}
%%%%%%%%%%%%%%%%%%%%%%%%%%%%%%%%%%%%%%%%%%%%%%%%%%%%%%%%%%%%%%%%%%%%%%

The interpretation of the absorption feature as a cyclotron scattering
line proposed by Rea et al.~(2003) was based on the following
criteria: 1) a Gaussian line gives a bad fit and does not reproduce
the asymmetrical shape of the observed feature; 2) the best fitting
model is the {\em XSPEC cyclabs} model; 3) the line strength is highly
phase dependent; 4) no atomic edges or absorption lines are known to
lie around 8.1\,keV (at least without assuming ad hoc shifts possibly
due to the high gravitational redshift or to Zeeman effects in such
strong magnetic field); 5) the relation between the line energy and
width agrees with that of cyclotron scattering features discovered in
other classes of sources (see Fig.\,5 in Rea et
al. 2003); 6) the magnetic field inferred from the line energy, either
in the case of an electron or proton cyclotron resonance, is
reasonably consistent with what is expected for a normal neutron star
($\sim10^{12}$\,G) or for a magnetar ($\sim10^{15}$\,G), both being
still open possibilities. Then, if this feature is real, all the above
points hint toward the cyclotron nature of the absorption feature at
8.1\,kev.

Keeping always in mind the possibility that the absorption line in the
\BSAX\, spectrum might be due to statistical fluctuations, in Rea et 
al.~(2005a) we discuss physical mechanisms which could be responsible
for the appearance of a transient cyclotron line in this source in the
context of the magnetar scenario.

\section{Discussion}
\label{diss}

\smallskip
{\bf 1. Intensity--Hardness correlation} 
\smallskip

The long-term evolution of the source flux and the spectral hardness
are shown in Fig.\ref{plot_all}, where different observations,
spanning nearly ten years, were collected. Comparing the two panels,
there seems to be a correlation between the photon index the source
X-ray flux. The spectrum became progressively harder as the flux rose
in correspondence of the two glitches and then softened as the
luminosity dropped, following the glitch recovery. This is suggestive
of a scenario in which the mechanism responsible for the glitches is
also at the basis of the enhanced emission and of the spectral
hardening.

The similarity of the second glitch of \rxj\, with the one discovered
during the bursting activity of \ea\ (Kaspi et al. 2003), after which
a similar exponential recovery was seen, suggests that bursts likely
occurred in \rxj\, as well, but the sparse observations did miss them,
as already suggested by Kaspi \& Gavriil (2003). Moreover, the
spectral parameters and the flux changes after the recovery of the
glitch strengthens this idea in comparison with what was reported
for the post-bursts fading of \ea\,(Woods et al. 2004).

This may be interpreted as the onset of a twist, which grew,
culminated in the glitches, and then decayed. A twisted external
field, in fact, is in an unstable magnetostatic equilibrium and
evolves towards a pure dipole field which represents the configuration
of minimal energy (see Rea et al.~2005a for a detailed interpretation).

The last Chandra and Swift observations show that the source is slowly
increasing again its flux and hardening its emission. Whether the
suggested correlation with the glitching and possibly bursting
activity holds, we expect the source to re-enter an active state
around 2006--2007. If confirmed for all the magnetars, the X-ray
monitoring might be an excellent tool to foresee the activity of
magnetars.

\smallskip
\noindent
{\bf 2. On \rxj\, IR counterpart}
\smallskip

Based on the deep VLT--NACO observation, we believe that the
identification of the IR counterpart of 1RXS J170849 --400910 is still an open issue,
mainly due to the very crowded region in which this source is located.
Note that the $2.5\sigma$ variability recently reported for source B
by Durant \& van Kerkwijk~(2006) does not seem to be a conclusive word
on the IR counterpart of this source, especially considering that
source B has IR magnitudes much brighter than all other AXPs. 

\smallskip
\noindent
{\bf 3. On the possible cyclotron line}
\smallskip

Following Protassov et al.~(2002) we re-addressed the issue of the
significance of the absorption line discovered by \BSAX. We end up
with a 99.68\% confidence level for its existence. The fact that the
source was not completely recovered by the second glitch at the time
the line appeared, make the correlation between line appearance,
glitching activity and flux enhancement possibly intriguing and might
suggests that the conditions for line formation were met at the
epoch of the \BSAX\ pointing. Future long X-ray observations are
needed in order to follow the source in its slow flux increase and
possibly re-detect the source in such state.

\begin{acknowledgements}

We thank Gordon Garmire for having observed \rxj\, with {\em Chandra}
within his Guarantee Time, and Cees Bassa for having recognised in the
different catalogues used for the astrometry, the reason for the shift
between our IR NACO field and the one reported by Durant \&
vanKerkwijk (2006). We also acknowledge F. Haberl, L. Kuiper and the anonymous referee 
for useful comments.

\end{acknowledgements}

\end{document}